\documentclass
[10pt,aps,prc,twocolumn,showpacs,showkeys,amsmath,floatfix,superscriptaddress]{revtex4-1}

\usepackage{graphicx}
\usepackage{mathptmx}                
\usepackage{dcolumn}                 
\usepackage{bm}                      

\usepackage{hyperref}
\usepackage{amssymb}
\usepackage{txfonts}
\usepackage{dcolumn}
\usepackage{slashed}
\usepackage{comment}
\usepackage{mathrsfs}
\usepackage{multirow}
\usepackage{epstopdf}
\usepackage{float}
\usepackage{color}
\usepackage{indentfirst}
\usepackage[section]{placeins}
\usepackage{CJK}

\def\be{\begin{equation}}
\def\ee{\end{equation}}
\def\bea{\begin{eqnarray}}
\def\eea{\end{eqnarray}}

\begin{document}

\title{Finite temperature pairing re-entrance in drip-line $^{48}$Ni nucleus}

\author{Mohamed Belabbas}
\affiliation{D\'epartement de Physique, Facult\'e des Sciences Exactes et Informatique, Hassiba Benbouali University of Chlef, P.O. Box 151, 02000, Ouled Fares, Chlef, Algeria }
\author{Jia Jie Li}
\affiliation{School of Nuclear Science and Technology, Lanzhou University, Lanzhou 730000, China}
\author{J\'er\^ome Margueron}
\affiliation{Institute for Nuclear Theory, University of Washington, Seattle, Washington 98195, USA}
\affiliation{Institut de Physique Nucl\'eaire de Lyon, CNRS/IN2P3, Universit\'e de Lyon, Universit\'e Claude Bernard Lyon 1, F-69622 Villeurbanne Cedex, France}

\begin{abstract}
Finite-temperature Hartree-Fock-Bogoliubov theory using Skyrme interactions and Relativistic Hartree-Fock effective Lagrangians, predicts $^{48}$Ni as being a possible candidate  for the finite temperature pairing re-entrance phenomenon. For this proton-drip-line nucleus, proton resonant states are expected to contribute substantially to pairing correlations and the two predicted critical temperatures are $T_{c1}\sim 0.08-0.2$~MeV and $T_{c2}\sim 0.7-0.9$~MeV. It is also shown that pairing re-entrance modifies the proton single particle energies around the Fermi level, as well as occupation numbers, and quasi-particle levels. The understanding of pairing re-entrance in $^{48}$Ni presently challenge our understanding of exotic matter under extreme conditions.
\end{abstract}

\date{\today}

\pacs{21.10.-k, 21.60.Jz}

\maketitle

\section{Introduction}
\label {s_intro}

Proton-rich nuclei provide interesting information on the strong interaction which is complementary to neutron-rich nuclei~\cite{Bla08,Bl08,Pf12}. 
Among them, $^{48}$Ni, with $28$ protons and $20$ neutrons, is one of the most proton-rich nuclei ever identified. 
It has been experimentally discovered in $1999$ at the SISSI/LISE3 facility of GANIL, where a lower limit for its half-life was found to be about 0.5 ms~\cite{Bl00}. This doubly magic nucleus located at the proton drip-line exhibits a remarkable stability with respect to external perturbation compared to neighboring nuclei. 
Owing to its doubly magic properties, $^{48}$Ni is also of particular interest because it is at the extreme limit of nuclear stability, where the nuclear forces are no longer able to bind all protons and neutrons together. Therefore, a possible decay mode of $^{48}$Ni was found to be the emission of two protons ($2p$ radioactivity). 
First indications of this new type of radioactivity has been found in an experiment at the SISSI/LISE3 facility of GANIL~\cite{Dos05} in 2004 and
confirmed at the National Superconducting Cyclotron Laboratory at MSU in 2011~\cite{Po12}.  

From the theoretical side, the ground-state properties of $^{48}$Ni and surrounding proton-rich nuclei have been widely studied within the nuclear shell-model~\cite{Or96}, the Hartree-Fock-Bogoliubov (HFB) theory~\cite{Na96}, and the Relativistic Hartree-Bogoliubov (RHB) theory~\cite{Vr98}. 
The realistic description of proton emitters requires further developments treating on an equal footing bound, resonant, and scattering states as well as the coupling to the decay channels~\cite{Bla08}. Let us mention some of the recent calculations in this direction: R-matrix~\cite{Grigorenko00,Grigorenko03}, three-body models~\cite{Brown03}, and shell model embedded in the continuum~\cite{Rotureau06}.
Schematic pairing approximations including resonance width have shown the interplay between pairing and resonant states for drip-line nuclei~\cite{Hasegawa03,Betan12}.
The effect on resonant states on pairing correlations was studied in the framework of BCS approximation, both for zero~\cite{Sa97,Sa00,Kr01} and finite  temperature~\cite{San00} and in the framework of HFB theory~\cite{De89,Ri80,Do84,Bulgac80,Be91,Meng98,Be99,Grasso01,Grasso02,Meng06,Zhang11,Pastore2013}, but essentially applied to neutron-rich nuclei.
In these papers, the important role of resonant states has been underlined, as nuclei get closer to the drip-lines.
More recently, a \textsl{pairing persistent} effect against temperature have been found~\cite{Mar12a,Jiajie}. Pairing persistence occurs if a finite amount of temperature could populate s.p. states above the Fermi level. The temperature should be less than the critical temperature, which is $\approx 1$~MeV in finite nuclei, impling that the excited states shall be less than $\approx 4$~MeV above the Fermi level. When the \textsl{pairing persistent} effect occurs, the critical temperature is found to be increased w.r.t the usual BCS estimation, $T_c\approx 0.57\Delta(T=0)$, where $\Delta(T=0)$ is the pairing gap at zero temperature.
If the ground state is unpaired, and a finite amount of temperature modifies the occupation of the s. p. orbitals enough to switch on pairing, then a very surprising behaviour called \textsl{pairing re-entrance} in the thermal equilibrium state could be observed.
In this case, the hole states around the Fermi level become unblocked at finite temperature and participate together with the excited states to the pairing correlations. 
This behaviour is going against the general rule that temperature destroys pairing, and as a consequence, this phenomenon may occur only at low temperature (below the critical temperature).

Pairing re-entrance phenomenon was first predicted for hot rotating nuclei by Kammuri~\cite{Ka64} and Morreto~\cite{Mo72}, and called \textsl{thermally assisted pairing} or \textsl{anomalous pairing}. Later it has also been predicted in odd-nuclei by Balian, Flocard and V\'{e}n\'{e}roni~\cite{Ba99} who have introduced the name \textsl{pairing re-entrance}. More recent studies of pairing re-entrance at finite temperature have been carried out for the rotational motion of nuclei~\cite{De10,Qu11,Sh05}, and the deuteron pairing channel in asymmetric matter~\cite{Se97}. At finite-temperature, pairing re-entrance in the equilibrium state was predicted for the first time in the neutron channel of the extremely neutron rich nuclei, $^{176-180}$Sn~\cite{Mar12a}. 

In this paper we investigate the finite temperature Hartree-Fock-Bogoliubov (FT-HFB) theory with Skyrme forces (FT-HFB)~\cite{Goodman81,Civitarese83,San00,Mar12b} and relativistic Lagrangians (FT-RHFB)~\cite{Jiajie}, which will be very briefly described in Sec.~\ref{sec:fthfb}. 
In Sec.~\ref{sec:results}, we discuss pairing re-entrance in $^{48}$Ni. 
Since this nucleus is located at the edge of present experimental possibilities, it has been produced only in a few numbers and it may hopefully be produced in larger amounts at the future experimental facilities, such as FAIR or FRIB.
Finally conclusions are presented in Sec.~\ref{sec:conclusions}.

\section{Finite temperature HFB}
\label{sec:fthfb}

For the sake of simplicity, we detail here some important FT-HFB equations with Skyme forces only. The more complex FT-RHFB equations can be found in our recent work~\cite{Jiajie}. Denoting $h_{T}(r)$ the thermal averaged mean field Hamiltonian and $\Delta _{T}(r)$ the thermal averaged pairing field, the radial FT-HFB equations read~\cite{Goodman81,Civitarese83,San00,Mar12b}, 
\begin{equation}
\left( 
\begin{array}{cc}
h_{T}(r)-\lambda & \Delta _{T}(r) \\ 
\Delta _{T}(r) & -h_{T}(r)+\lambda%
\end{array}%
\right) \left( 
\begin{array}{c}
U_{i}(r) \\ 
V_{i}(r)%
\end{array}%
\right) =E_{i}\left( 
\begin{array}{c}
U_{i}(r) \\ 
V_{i}(r)%
\end{array}%
\right)  \label{1}
\end{equation}
where $E_{i}$ stands for the positive quasi-particle energy eigenvalue, $U_{i}$ and 
$V_{i}$ are the components of the radial FT-HFB wave function and $\lambda$
is the Fermi energy associated to the particle conservation equation. 
For the zero-range Skyrme forces, the FT-HFB Hamiltonian and pairing field
can be written in terms of the particle density,
\begin{eqnarray}
\rho(r)&=&\frac{1}{4\pi }\sum_{i}(2J_{i}+1)\left[ V_{i}^{\ast
}(r)V_{i}(r)(1-f_{i})+U_{i}^{\ast }(r)U_{i}(r)f_{i}\right] , \nonumber \\ \label{2}
\end{eqnarray}
where $f_{i}=\left[ 1+\exp (E_{i}/k_{B}T)\right] ^{-1}$ is the thermal occupation probability of quasi-particle states with angular momentum $J_{i}$, $k_{B}$ is the Boltzmann constant and $T$ is the temperature.  The FT-HFB Hamiltonian depends also on the thermal spin- and kinetic-densities which are defined, for instance, in Ref.~\cite{Mar12b}. In Eq.(\ref{2}), the summation is going over the whole quasi-particle spectrum. The thermal average pairing field is calculated with a density dependent contact force of the following form~\cite{Be91}: 
\begin{equation}
V(r-r^{\prime })=V_{0}\left[ 1-\eta \left( \dfrac{\rho (r)}{\rho_{sat}}%
\right) ^{\alpha }\right] \delta (r-r^{\prime })  \label{5}
\end{equation}%
where $\rho(r)$ is the density ($\rho_{sat}=0.16$~fm$^{-3}$) and  $V_{0}$ is the strength of the force. We have considered a mixed surface/volume pairing interaction~\cite{Bertsch2009,Yamagami2012} by fixing the two other parameters $\eta$ and $\alpha$ to the following values $\eta=0.7$ and $\alpha=0.45$. With this force the thermal averaged pairing field is local and is given by,
\begin{eqnarray}
\Delta _{T}(r)&=&
V_{eff}[\rho (r)] \kappa _{T}(r)
\label{6}
\end{eqnarray}%
where $\kappa _{T}(r)$ is the thermal averaged pairing tensor given by,
\begin{equation}
\kappa _{T}(r)=\frac{1}{4\pi }\sum_{i}(2J_{i}+1)U_{i,q}^{\ast
}(r)V_{i}(r)(1-2f_{i})  \label{7}.
\end{equation}%

\begin{figure}[t]
\includegraphics[width=1.0\columnwidth]{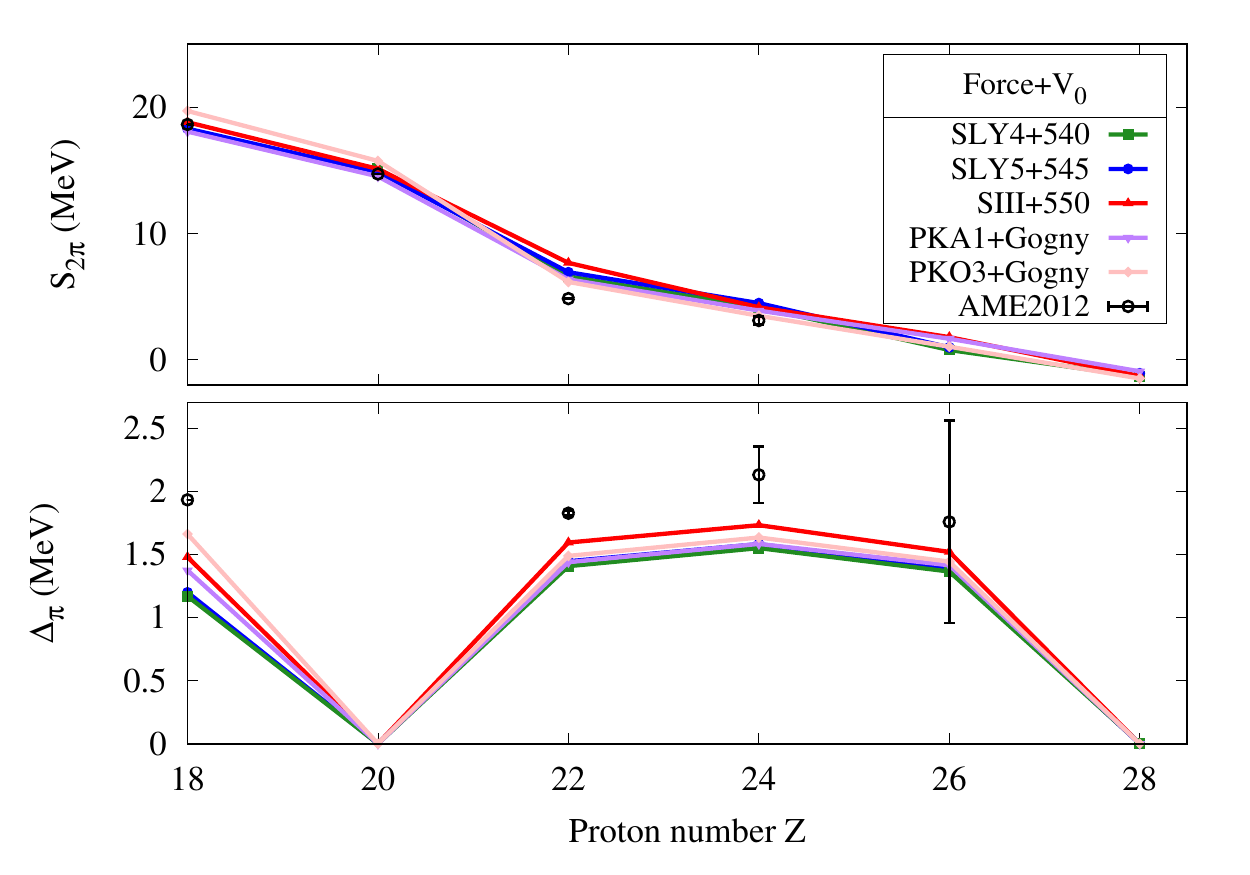}
\vspace{0.3cm}
\caption{(Color online) Two-proton separation energies (upper panel) and proton pairing gaps (lower panel) versus the number of protons for $N=20$ isotones obtained within the FTHFB model at zero temperature, for SLY4-5~\cite{sly} and SIII~\cite{siii} Skyrme forces and PKA1~\cite{Long07} and PKO3~\cite{Long08} RHFB Lagrangians. The experimental data are deduced from AME2012 mass table and the experimental three-points formula is used for the comparison to the proton pairing gap.}
\label{fig1}
\end{figure}

The thermal average pairing gap is obtained from the thermal average pairing field $\Delta _{T}(r)$ and the thermal pairing tensor $\kappa _{T}(r)$ solutions of the finite-temperature HFB model as~\cite{Sa04,Mar12b},
\begin{equation}
\Delta \equiv \frac{\int d^{3}r\Delta _{T}(r)\kappa _{T}(r)}{\int
d^{3}r\kappa _{T}(r)}  \label{8} .
\end{equation}
In practice, the self-consistent FT-HFB equations (1) are solved by iterations, fixing at each iteration the chemical potential $\lambda $ and the pairing field $\Delta _{T}(r)$, until the convergence of the  total energy. The wave functions $U_i$ and $V_i$ are solved in coordinate space using the Numerov method with Dirichlet asymptotic boundary conditions~\cite{Do84}. The size of the box is fixed to be 30~fm and the step in coordinate space is 0.2~fm. 
The cut-off is fixed to be 60~MeV and the maximal value of the angular momentum considered here is $J_{max}=30$.
We have checked the stability of our results against these parameters and found convergence, see for instance Fig.~4 of Ref.~\cite{Mar12a} for more details. 
In the present work, continuum states are represented by the positive-energy states of the box. In doing so, we neglect the effect of the resonance state widths which is expected to reduce the pairing correlations~\cite{Meng98,Sa00,San00,Grasso01,Grasso02,Meng06,Zhang11}.
In the present calculation, we carefully adjust the pairing interaction to the two-neutron separation energy of near-by nuclei and check the proton pairing gap against the three-point formula, see Fig.~\ref{fig1}, in order to minimise the error induced by our approximation for the continuum states.

We have searched for the occurrence of pairing re-entrance using various nuclear interactions and running over magic nuclei and their neighbors. In more details, we have performed systematical calculations with various effective Skyrme interactions such SLY4-5~\cite{sly}, SKMS~\cite{skms}, SKI1-5~\cite{ski}, SGII~\cite{sgii}, SIII~\cite{siii}, and RATP~\cite{ratp}, for many nuclei, namely, $^{14}$C, $^{12,16,22,24}$O, $^{30}$Ne, $^{34,40,48,58}$Ca, $^{34}$Si, $^{38}$Ar, $^{48,56,60,66,76}$Ni, $^{82}$Ge, $^{90}$Zr, $^{100,120,132}$Sn, $^{140}$Ce, $^{146,190}$Gd and $^{176,208}$Pb. These nuclei are semi-magic or doubly magic and they are located at or close to the drip-lines, which is a condition for pairing re-entrance~\cite{Mar12a}. We have focused on nuclei which have already been produced at nuclear facilities. From these extensive studies, we have found that only the doubly magic neutron-deficient nucleus $^{48}$Ni may manifest the re-entrance phenomenon in its thermally equilibrated state. 

In the case of $^{48}$Ni and surrounding nuclei, the value of the pairing strength $V_0$ have been determined for each Skyrme interactions such as to the experimental two-proton separation energies $S_{2p}$ and the proton pairing gaps $\Delta_p$ for $N=20$ determined from the three-point formula~\cite{Sa98,Changizi15,Afanasjev15}, where the experimental nuclear masses are provided by the AME2012 mass table~\cite{ame2012}, are in overall agreement within the known differences between these quantities~\cite{Bertulani2009}. The comparison between the model predictions for $S_{2p}$ and $\Delta_p$, and the experimental data is shown in Fig.~\ref{fig1}. The values of $V_0$ (in MeV.fm$^3$) are given in the legend of Fig.~\ref{fig1} for the Skyrme interactions SLY4~\cite{sly}, SLY5~\cite{sly} and SIII~\cite{siii}. The two other parameters $\eta$ and $\alpha$ are not modified. 
For the RHFB Lagrangian PKA1~\cite{Long07} and PKO3~\cite{Long08} the pairing interaction derived from the Gogny D1S~\cite{Berger84} finite-range interaction, see Ref.~\cite{Jiajie,Long2010a,Long2010b} for more details. 
There is a good agreement between the model predictions and the experimental data, given the experimental and systematical uncertainties. Notice that only the experimental uncertainties for mass measurements are represented in Fig.~\ref{fig1}. The evaluation of the systematical uncertainties, especially for $\Delta_p$ is difficult to estimate, but it is expected to be of the order of a few 100~keV~\cite{Sa98,Changizi15,Afanasjev15}.

\begin{figure}[t]
\includegraphics[angle=-90,width=1.05\columnwidth]{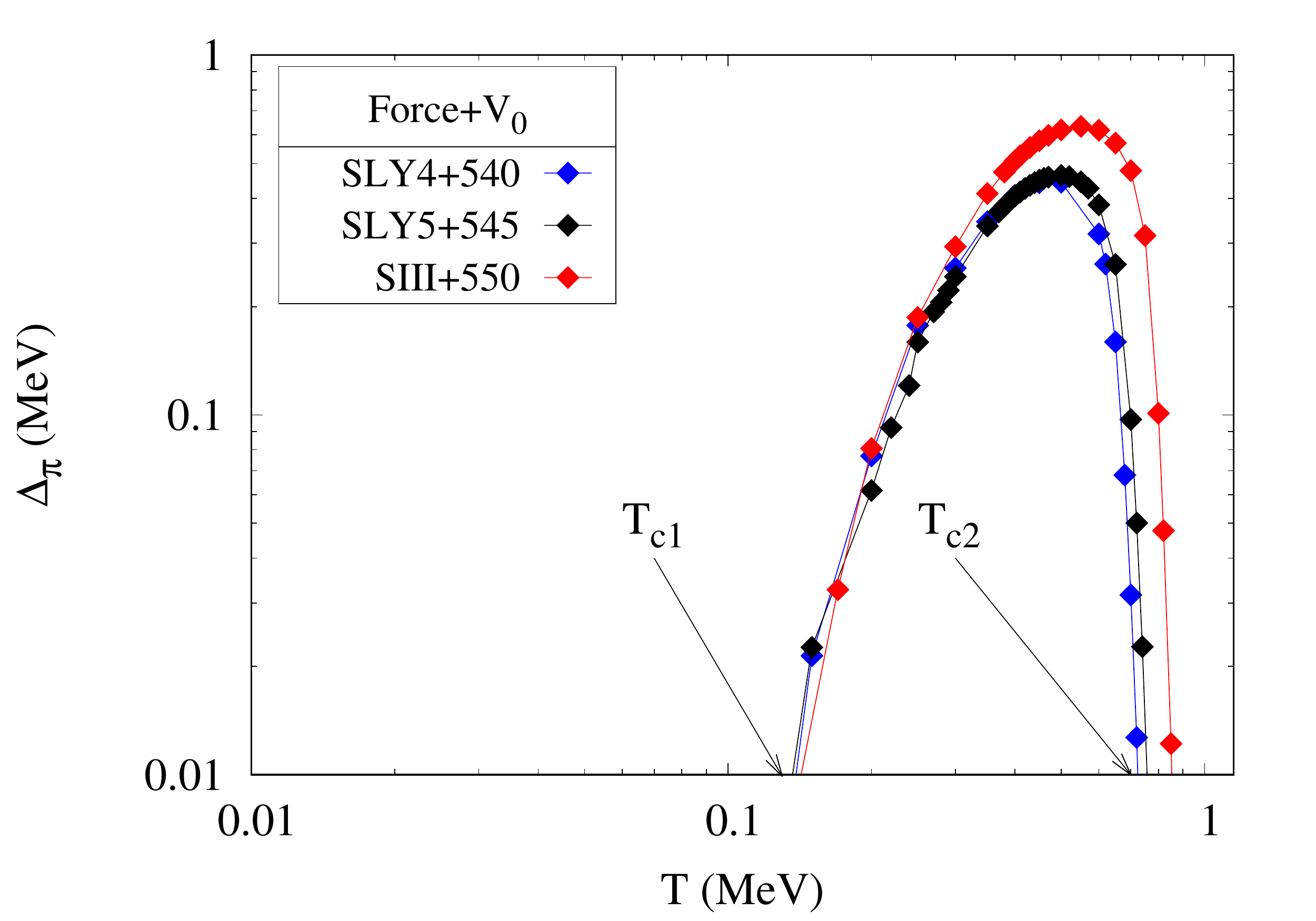}
\caption{(Color online) Temperature-averaged proton pairing gap versus temperature for $^{48}$Ni nuclei based on SLY4-5 and SIII Skyrme
interactions. See text for more details.}
\label{fig2}
\end{figure}

\section{Results and discussion}
\label{sec:results}

The predictions for the temperature-averaged proton pairing gaps $\Delta_p$ are shown in Fig.~\ref{fig2} for the same interactions as in Fig.~\ref{fig1}. The pairing re-entrance phenomenon is predicted for SLY5, SLY4, and SIII Skyrme forces with critical temperatures $T_{c1}\sim 0.1-0.2$~MeV and $T_{c2}\sim 0.7-0.8$~MeV. These critical temperatures correspond to the low- and high-temperature boundaries of the pairing re-entrance domain. Out of this domain, matter is predicted to be in its normal phase where pairing is quenched. Notice that the high-temperature boundary $T_{c2}$ is below or about the same as the single critical temperature in ordinary nuclei~$\sim 1$~MeV~\cite{Goodman81,Civitarese83,San00,Mar12a,Jiajie}. Other Skyrme forces considered here do not predict the re-entrance phenomenons for reasons that we detail hereafter.

\begin{figure}[t]
\includegraphics[width=0.95\columnwidth]{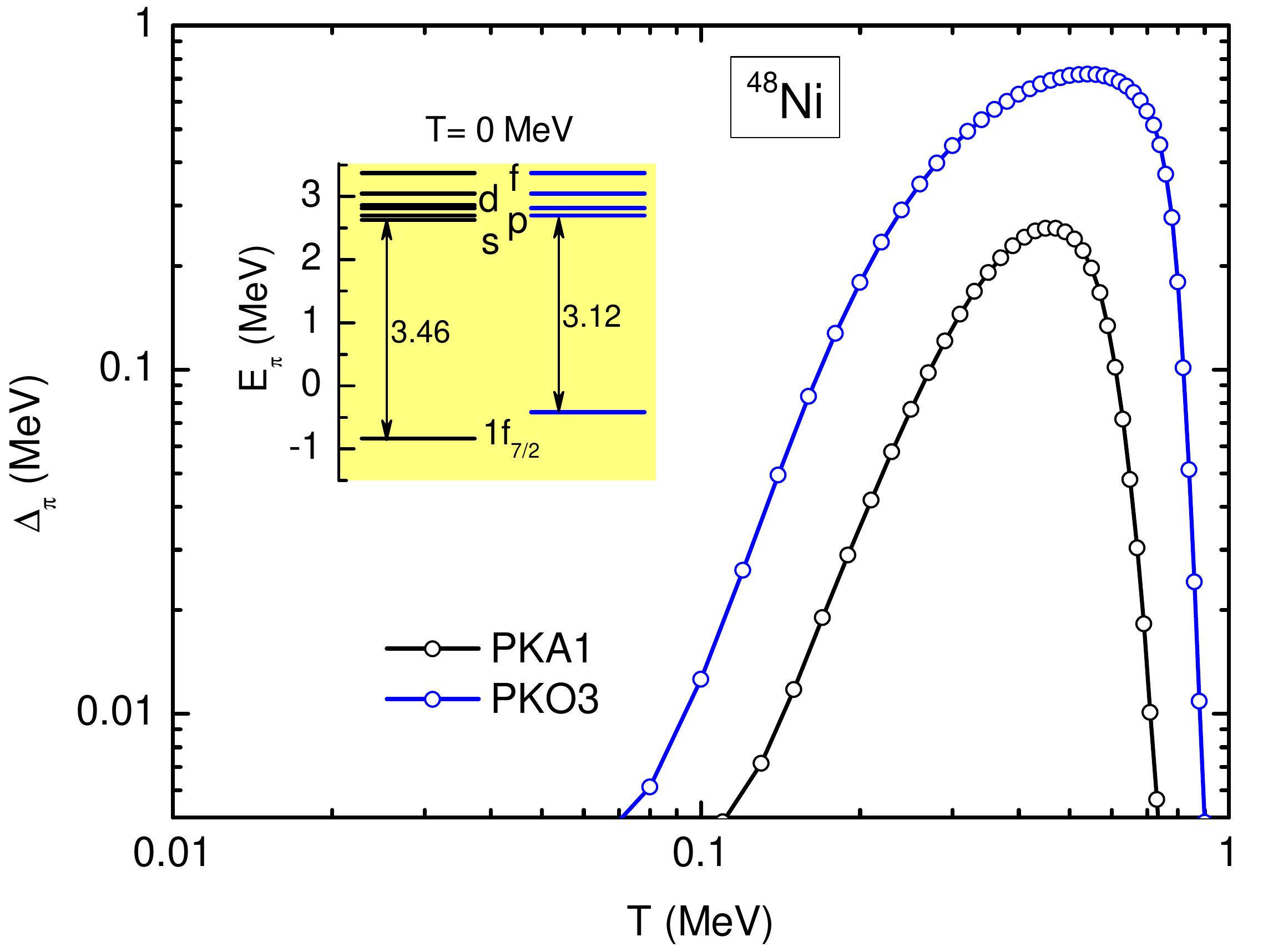}
\caption{(Color online)  Prediction for the proton pairing gap in Ni$^{48}$ as function of the temperature based on 
RHF PKA1~\cite{Long07} and PKO3~\cite{Long08} effective Lagrangians.  See text for more details.}
\label{fig4}
\end{figure}

Finite temperature pairing re-entrance is also predicted by other interaction models. For instance, the prediction from relativistic Hartree-Fock-Bogoliubov effective Lagrangians PKA1~\cite{Long07} and PKO3~\cite{Long08} are shown in Fig.~\ref{fig4}.
These Lagrangians are considered to be among the best ones presently existing since they are based on the exchange of $\sigma$, $\omega$, $\rho$ and $\pi$ meson and consistently include the Fock exchange term.
The predictions of PKA1 and PKO3 for the critical temperatures $T_{c1}$ and $T_{c2}$ are quite similar to the one based on SLy4, SLY5 and SIII, see Fig.~\ref{fig2}: they predict a domain of temperature for the re-entrance phenomenon going from 0.08-0.1~MeV up to 0.7-0.9~MeV. The strength of the pairing gap at maximum varies from one interaction to another. The single particle energies are given in the inset of Fig.~\ref{fig4} for PKA1 (left) and PKO3 (right), predicting a proton gap of the order of 3.12~MeV for PKO3 and 3.46~MeV for PKA1. It is quite logical, since the s.p. gap is slightly smaller for PKO3 compared to PKA1, that the pairing re-entrance domain as well as the value of the proton pairing gap are bigger for PKO3 compared to PKA1. Notice that the relative low-energy s.p. gap of the $Z=28$ shell favors the appearance of the re-entrance pairing correlation.

The structure of the single-particle states around the Fermi energy provides a good understanding of the theoretical results for pairing re-entrance. As nuclei get closer to the drip-lines, the coupling to the continuum becomes more and more important, and continuum resonant states may play an important role if they are located at low energy~\cite{Do84,Mar12a,Pastore2013,Jiajie}.
For pairing re-entrance, it is important that these resonant states are sufficiently high (above about 2~MeV) such that the ground-state is unpaired, but at the same time, it shall be sufficiently low (below about 4~MeV) to be populated by low-temperature thermal excitation~\cite{Mar12a,Jiajie}.
Notice that the energy boundaries given here are only illustrative and could not be used to predict if pairing re-entrance occurs or not. 
These boundaries change with the pairing strength, which is known to change through the nuclear chart.
The critical temperature $T_{c1}$ depends on the position of the resonant state, and the closer it is to the Fermi energy, the lower is $T_{c1}$.
The other critical temperature $T_{c2}$ is limited to a value which is about 1~MeV, as  for the usual critical temperature in ordinary paired nuclei~\cite{Goodman81,Civitarese83,San00,Mar12a,Jiajie}.
The quenching mechanism is indeed the same in the re-entrance case and in ordinary paired nuclei:
the single particle thermal excitation breaks the Cooper pairs, since the cost in kinetic energy of having particles well above the Fermi energy is not anymore compensated by the gain in forming Cooper pairs.
Since the quenching mechanism is the same for ordinary paired nuclei and for pairing re-entrance, the critical temperature $T_{c2}$ is also limited to values around about 1~MeV.

\begin{figure}[t]
\includegraphics[width=1.0\columnwidth]{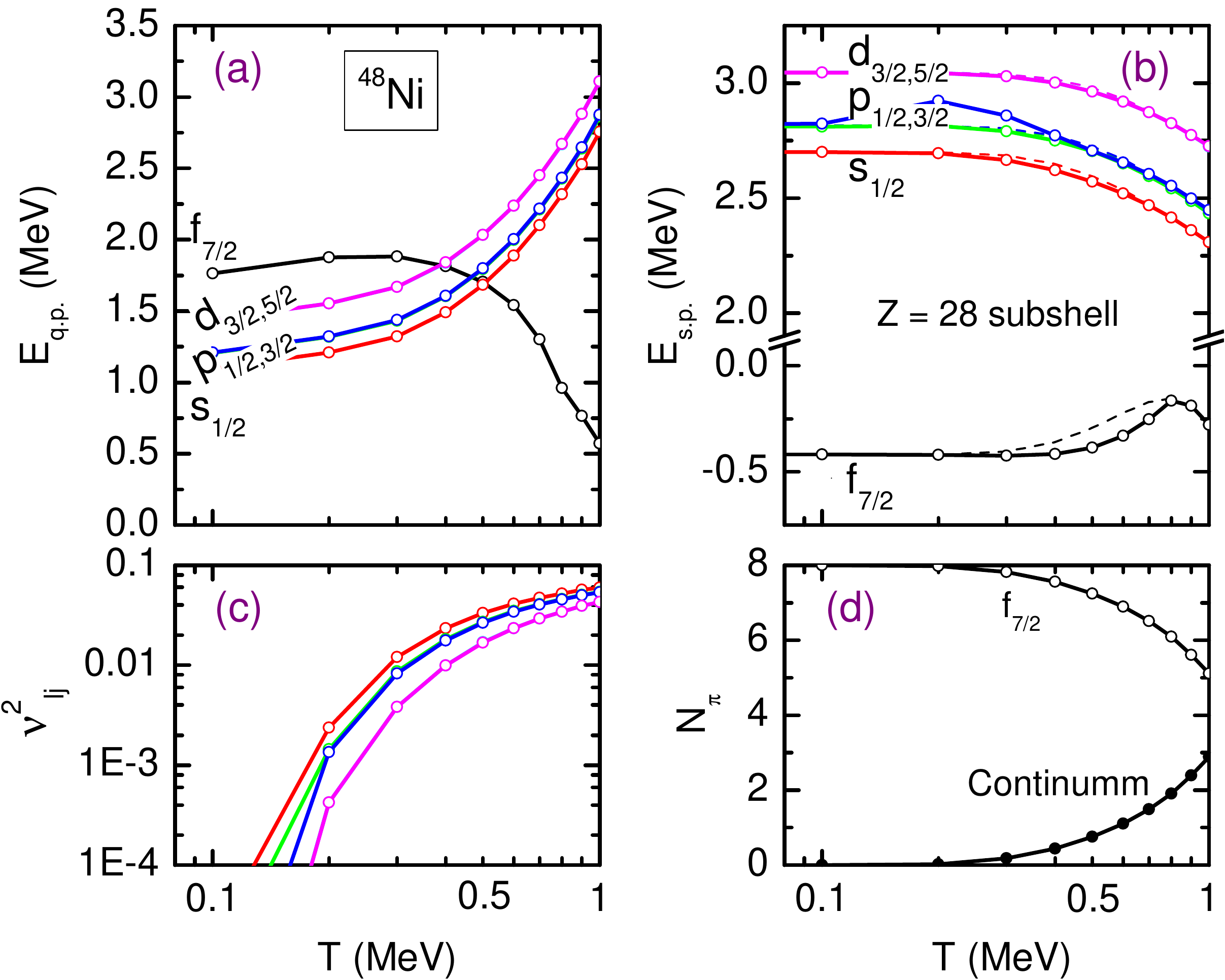}
\caption{(Color online) (a) Temperature evolution of the proton quasi-particle energies corresponding to the states around the Fermi energy: 1f7/2 (hole), and s1/2, p1/2-3/2, and d3/2-5/2 (particles). (b) Temperature evolution of the same proton states in the canonical basis. (c) Occupation numbers of the particle states function of the temperature. (d) Occupation numbers of the 1f7/2 hole state as function of the temperature. In the (d) panel, we also represent the sum of the occupation probabilities (called continuum) for the particle states shown in panel (c).
These are results of FT-RHFB with PKO3 effective Lagrangian and the Gogny pairing force D1S.}
\label{fig3}
\end{figure}

In order to understand the behaviour of the nuclear structure of $^{48}$Ni in the region of re-entrance, we now analyze results obtained from FT-RHFB with PKO3 effective Lagrangian and the Gogny pairing force D1S, see Ref.~\cite{Jiajie} for more details on the theory side. The evolution of the proton properties around the Fermi energy with respect to the temperature is shown in Fig.~\ref{fig3} (top panel) for temperatures between 0.08 and 1~MeV. The quasi-particle states (panel a) are increasing function of the temperature for particles and decreasing for holes. Assuming that the quasi-particle energy is related to the s.p. energies (shown in panel b) by the following relation
$E_{qp}=\sqrt{(e_{s.p.}-\mu)^2+\Delta^2}$, and knowing that the chemical potential $\mu(T)$ is a decreasing function of the temperature, it can be understood that for constant $e_{s.p.}$ and $\Delta$, the quasi-particle energy decrease for hole states and increase for particle states. The s.p. energies $e_{s.p.}$ shown on panel (b) of Fig.~\ref{fig3} are almost constant up to $T\sim 0.5$~MeV and change by about 200~keV at $T\sim 1$~MeV. 
The thin dashed lines in Fig.~\ref{fig3}(b) show the T-dependence of s.p. states in the absence of pairing correlation (pairing interaction has been numerically quenched). The impact of pairing correlations can therefore be estimated comparing the thin dashed and the solid lines. Pairing correlations tend to stabilize the T-dependence of f7/2 state up to $T\sim 0.5$~MeV, while the other states are almost unmodified.
The effect of temperature is to populate particle states, which are s1/2, p1/2-3/2, and d3/2-5/2 states, while depopulating hole states, such as the f7/2 state.
The effect of the temperature on changing the occupation numbers is shown in panels (c) and (d) of Fig.~\ref{fig3}.
Panel (c) shows the increasing occupation numbers of the particle states as function of temperature, and panel (d) the sum of particle states, labeled Continuum,  against the occupation number of the f7/2 state. There is an almost perfect symmetry between the occupation numbers of f7/2 states and the continuum states, showing that the main states playing an important role in the re-entrance phenomenon are the f7/2, s1/2, p1/2-3/2, and d3/2-5/2 states.
The other nuclear interactions shown in Figs.~\ref{fig2}-\ref{fig3} predict similar qualitative behaviour of the quasi-particle properties.
Notice that changing the size of the box has a very marginal impact, as it has already been shown in Ref.~\cite{Mar12a}.

On the experimental side, the observation of the pairing re-entrance phenomenon is very challenging and requires the production of a large amount of $^{48}$Ni, which is yet impossible.
One might think in a first step to better investigate the position of the resonant states in the continuum through one-proton transfer reactions. 
While not being a direct probe of the pairing re-entrance phenomenon, such a preliminary experimental investigation would test the necessary condition to make finite temperature pairing re-entrance possible.
Two-proton transfer could also be considered, where thermal excitation may be induced by highly charge incident particle.
Ultimately, in a farther future when a very large amount of $^{48}$Ni will be available, the study of hot giant resonances in $^{48}$Ni may provide a clear signal
to probe the thermal pairing re-entrance phenomenon.

\section{Conclusions}
\label{sec:conclusions}

In summary, based on FT-HFB approach, we found that $^{48}$Ni may be the only nucleus presently synthesised where the finite temperature re-entrance phenomenon in the thermal equilibrium state may occur. 
This prediction has to be tested against improved nuclear modelling.
The domain of temperature where this phenomenon could occur, as well as the size of the proton pairing gap, still depends on the detailed s.p. level structure which varies from one interaction to another. 
The treatment of the continuum states in the present FT-HFB approach shall be improved in the future, and important questions related to the effect of particle number restauration or additional correlations shall also be investigated. The present work is however the first one suggesting that $^{48}$Ni may be reentrant at finite temperature. 
The present nuclear theories, as well as the experimental facilities, are still far from being able to provide a clear understanding of exotic matter under extreme condition, and our prediction challenges them both further.

\end{document}